\documentstyle[amssymb,prl,aps,psfig]{revtex}

\begin{document}
\twocolumn
\title{Anomalous behaviour of S/N mesoscopic structures near $T_c$.}
\author{S.Shapira$^{+}$, E.H.Linfield$^{+}$, C.J. Lambert$^{\ast }$, R. Seviour $%
^{\ast }$, A.F. Volkov$^{\ast \dagger }$ and A.V. Zaitsev $^{\dagger }$}
\address{$^{+}$Cavendish Laboratory, University of Cambridge, \\
Madingley Road CB3 0HE,\\
UK \\
$^*$ School of Physics and Chemistry,\\
Lancaster University, Lancaster LA1 4YB, U.K.\\
$^{\dagger}$Institute of Radioengineering and Electronics of the Russian \\
Academy of Sciencies, Mokhovaya str.11, Moscow 103907, Russia.}
\date{\today}
\maketitle

\begin{abstract}
We observe a maximum in the conductance of Aluminum / n-GaAs junctions at
temperatures 20 mK lower than the superconducting transition temperature.
This is the first observation of a peak in the conductance near the
superconducting transition in superconducting normal junctions. To
accommodate this effect we calculate the full temperature dependence of the
conductance of these structures, invoking quasiclassical Green's functions
in the diffusive limit. In addition to the well-known low temperature peak
at temperatures on the order of the Thouless energy, we find a maximum near
the critical temperature. This peak has the same origin as the subgap
conductance observed in superconducting-normal junctions at low
temperatures. Its calculated magnitude and position are in reasonable{\bf \ }%
agreement with our experimental observation.
\end{abstract}

\smallskip

Pacs numbers: 72.10.Fk, 72.15.Gd \smallskip

The study of transport properties of mesoscopic normal superconducting (N/S)
structures has revealed a number of novel features \cite{r1,r2}, including
long-range, phase-coherent transport, disorder-enhanced subgap conductance 
\cite{r3} and a non-monotononic dependence of the conductance on temperature
($T$) and voltage ($V$). Such a non-monotonic dependence was first predicted
in [4] for a short point N/S contact of length $L$ satisfying, $L^{2}<<\hbar
D/\Delta (0),$ where $D$ is the diffusion coefficient and $\Delta (0)$ is
the energy gap of the superconductor at zero temperature. In Ref.\cite{r4}
it was shown that $G$, as a function of $T$ or $V$, increases from $G_{n}$ ($%
G_{n}$ is the conductance in the normal state) at $T=0$ and $V=0$, reaches a
maximum at $T$ or $eV$ of order $\Delta (T)$ and then decreases to $G_{n}$
as $T$ or $V$ are increased further. In the opposite limit, 
\begin{equation}
\epsilon _{L}<<\Delta (0),  \label{eq1}
\end{equation}
where $\epsilon _{L}=\hbar D/L^{2}$ is the Thouless energy, the conductance
reaches a maximum when $T$ or when $eV_{m}$ is of order $\epsilon _{L}$ \cite
{r5,r6,r7,r8,r9}. These predictions have stimulated many experimental
studies of the sub-gap conductance of S/N systems and this low temperature
maximum has been observed in both short and long S/N structures \cite
{r10,r11,r12,r13}.

In Refs. \cite{r14,r15} it was predicted that besides the maximum $\epsilon
_{L}$, which is an energy scale distinct from $\Delta $, a new maximum in
the conductance may appear near $T_{c}$ when $\Delta (T)$ is of order $%
\epsilon _{L}$\cite{ftn1}. This new maximum arises from an interplay between
the two hitherto decoupled energy scales, $\epsilon _{L}$ and $\Delta $.{\bf %
\ }Thus one can probe both the energy gap $\Delta (T)$\ in the
superconductor and a pseudogap induced in the normal region, at an easily
accessible (viz. high) temperature range. The pseudogap dependence on
various parameters such as the magnetic field and Coulomb interaction can
then be tested against theoretical predictions \cite{r17,r18,r19}. Until now
however, this maximum has remained experimentally elusive.

We present here the first experimental observation of a maximum in the
conductance of S/N and S/N/S structures in the vicinity of $T_{c}$. The
structures measured are of a coplanar geometry (Top Inset in Fig. 2; see
also Fig.1(B)) in which the separation between the superconducting contacts
is larger than the coherence length $\xi =\sqrt{\hbar D/2\pi k_{B}T}$ ( $D$
is the diffusivity) in the GaAs layer. Our results for this structure are
applicable in particular to the limit described by Eq. (\ref{eq1}), although
the parameters and geometry are different than those used in \cite{r14,r15}.
For the first time we identify a significant conductance maximum near $T_{c}$%
, which occurs when $\Delta (T)$ is roughly equal to $\hbar D/d^{2}$ divided
by the ratio of the Al/GaAs interface resistance to the resistance through a
vertical portion of the conducting GaAs layer the thickness of which is $d$.

The junctions were made of 200 nm Al deposited in-situ in a molecular beam
epitaxy chamber, on heavily doped GaAs (the thickness of \ the heavily doped
region is{\bf \ }$d=$ 150 nm) \cite{ftn2}. Conducting bars of width $%
W=7,9,25\mu $m were then patterned by an acid etch. Finally an aluminium
strip of length $L=0.9-7\mu $m was removed by a second acid etch, thus
defining the two contacts (Top left Inset of Figure $2$). The conductance
was then measured via a standard four terminal low frequency (77Hz) ac
lockin technique. The data presented in Fig.2 is the conductance of a
junction patterned on a bar 7 microns wide with a contact separation of 0.9
microns. The sample is characterized by the following parameters \cite{ftn2}%
: the sheet resistances of the normal aluminium layer and the doped
semiconducting layer (at $T=1.5K$) are 0.02 $\Omega $ and 90 $\Omega $ per
square respectively. The interface Al/GaAs-n resistance is $R_{S/N}\cong
10^{-7}\Omega \cdot $cm$^{2}$, the average carrier density in GaAs-n$^{\text{%
+}}$ is 3.6$\times $10$^{\text{18}}$ cm$^{-3}$ (it varies \ in that layer
between 3$\times $10$^{\text{18}}$ - 5$\times $10$^{\text{18}}$cm$^{-3}$)
yielding the diffusion coefficient $D=110\cdot $cm$^{2}/$s. Thus the
coherence length $\xi (T)$ is 0.2(0.1) microns at T = 0.3(1.3) K, shorter
than the contact separation.

As Fig.\ref{fig2} shows, at the superconducting transition, the conductance
increases, reaching a maximum when the temperature is 20 mK colder than the
point at which\ the conductance was first affected \cite{ftn3}. The trend is
then reversed as the temperature is further decreased. Note that a decrease
in the conductance (increase in the resistance) was earlier observed by
Kleinsasser and Kastalsky \cite{r3} on the Nb/InGaAs structure. We have
observed the effect in 9 samples including both S/N and S/N/S junctions with
varying contact separations ($0.9\mu $m$<L<7\mu $m) and the bar width ($%
W=7,19,25\mu $m). The magnitude of the conductance enhancement at the
transition divided by the contact width $W$ is of the same magnitude for all
samples, thus confirming that it is a property of one S/N junction.

\begin{figure}
\centerline{\psfig{figure=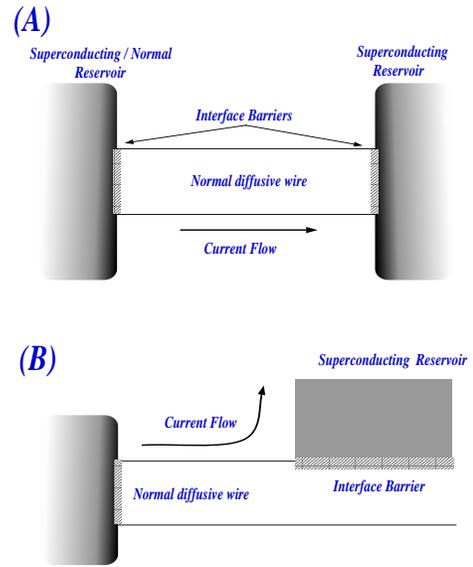,width=7cm,height=8cm}}
\caption{ The two structures considered in this paper. (A) S/N/S and N'/N/S
system, 1-D normal diffusive wire in contact with a supercoducting reservoir
and a normal/superconducting reservoir. (B) The coplanar structure.}
\label{fig1}
\end{figure}

\begin{figure}
\centerline{\psfig{figure=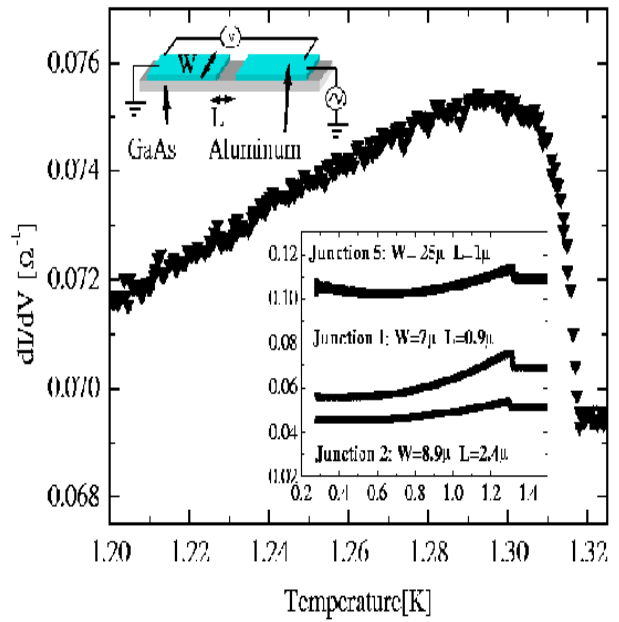,width=10cm,height=10cm}}
\caption{The conductance of junction 1 in the vicinity of $T_{c}$ Right
Inset: The conductance at a wide temperature range for three junctions,
including 1 Top Left Diagram: The junction structure.}
\label{fig2}
\end{figure}

To calculate the conductance variation of the structures shown in Fig. \ref
{fig1} (A,B). we assume the diffusive limit and exploit a quasiclassical
Green's function technique which has been well developed for studying
transport in S/N structures (see Ref.\cite{r2} and references therein). The
normalized conductance variation of the N'/N/S structure (N' is the normal
reserviour), $\delta S=(G_{s}-G_{n})/G_{n}$, is given by (see Ref.\cite{r22}%
), 
\begin{equation}
\delta S=\beta \int d\epsilon F_{V}^{\prime }\frac{1-<m^{-1}>+r_{N}-\frac{%
r_{N}}{\nu (0)}+r_{S}-\frac{r_{S}}{\nu _{S}\nu (L)+b(L)}}{<m^{-1}>+\frac{%
r_{N}}{\nu (0)}+\frac{r_{S}}{\nu _{S}\nu (L)+b(L)}},  \label{eq2}
\end{equation}
Here $F_{V}^{\prime }=1/2[cosh^{-2}((\epsilon +eV)\beta
)+cosh^{-2}((\epsilon -eV)\beta )]$ is the derivative of the distribution
function in the N' reservoir (we set the electric potential in S equal to
zero), $\beta =(2T)^{-1}$, $r_{N,S}$ are the ratios of the N'/N and N/S
interface resistances to the resistance of the N layer. The denominator
represents contributions to the total resistance at a given energy $\epsilon 
$ (or at a given voltage at zero temperature). The first term in the
denominator $<m^{-1}>$ describes the spatially averaged spectral resistance
of the N layer, which is decreased due to the proximity effect. The second
and third terms in the denominator are the N'/N and N/S interface spectral
resistances. The terms $\nu $ and $\nu _{S}$ representing the DOS in the N
layer and the S reservoir lead to an increase of the N'/N and N/S interface
resistances. The function $b$ represents a factor contributing to the subgap
conductance, which may lead to a decrease of the S/N interface resistance.
All the functions in Eq. (\ref{eq2}) are expressed in terms of the retarded
Green's functions $G^{R}$ and $F^{R}$ in the N film as 
\begin{equation}
\begin{array}{c}
<m^{-1}>=L^{-1}\int_{0}^{L}dx\cdot m^{-1}(\epsilon ,x), \\ 
m(\epsilon ,x)=[1+|G^{R}(\epsilon ,x)|^{2}+|F^{R}(\epsilon ,x)|^{2}]/2, \\ 
\nu (\epsilon ,x)=\Re \left( G^{R}(\epsilon ,x)\right) \\ 
\nu _{S}=\Re \left( G_{s}^{R}(\epsilon ,x)\right) =\Re \left( (\epsilon
+i\Gamma )/\sqrt{(\epsilon +i\Gamma )^{2}-\Delta ^{2}}\right) \\ 
b(\epsilon ,x)=\Im F^{R}(\epsilon ,x)\cdot \Im F_{S}^{R}(\epsilon )
\end{array}
\label{eq3}
\end{equation}
In order to calculate the conductance variation $\delta S$ given by Eq.(\ref
{eq2}), we need to find the Green's functions $G^{R}$ and $F^{R}$. These
functions obey the Usadel equation which after parametrization $G^{R}=\cosh
u^{R}$, $F^{R}=\sinh u^{R}$ acquires the form, 
\begin{equation}
\partial _{xx}^{2}u^{R}-(k_{\epsilon }^{R})^{2}\sinh u^{R}=0.  \label{eq4}
\end{equation}
where $k_{\epsilon }^{R}=\sqrt{(-2i\epsilon )/D}$. Eq. (\ref{eq4}) is
complemented by boundary conditions (see for example [2]).

Eq. (\ref{eq4}) can be solved analytically in two cases. The simplest case
corresponds to a weak proximity effect when the Usadel equation can be
linearised. This can be done if the condition $r_{S}>>r_{N}$ is satisfied.
The second case corresponds to a short structure \cite{r14}. We are
interested in energies of the order $\epsilon _{S}$\ (see below) which are
much smaller than the critical temperature $T_{c}.$\ Hence the relevant
length scale in Eq.(\ref{eq4}) $L_{S}=\sqrt{\hbar D/\epsilon _{S}}$\ is
longer than the characteristic length for the Josephson effect $\xi =\sqrt{%
\hbar D/2\pi T_{c}}$. In a short contact (that is, $L<L_{S}$), the solution
of Eq. (\ref{eq4}) is almost uniform along the layer for energies of order $%
\epsilon _{s}$, and has the form \cite{r14}, 
\begin{equation}
G^{R}=\epsilon ^{R}/\xi ^{R},\ F^{R}=\epsilon _{S}/\xi ^{R}  \label{eq6}
\end{equation}
where $\epsilon ^{R}=\epsilon (\zeta ^{R}+\epsilon _{S}/\Delta )+i\zeta
^{R}\epsilon _{N}$ , $\epsilon _{S,N}=\epsilon _{L}/2r_{S,N}$, $\zeta ^{R}=%
\sqrt{1-(\epsilon +i\cdot \Gamma )^{2}/\Delta ^{2}\text{,}}$ $\xi ^{R}=\sqrt{%
(\epsilon ^{R})^{2}-\epsilon _{s}^{2}},\Gamma $ is the damping in the energy
spectrum of the superconductor S, and $\epsilon _{L}=\hbar D/L^{2}$ is the
Thouless energy. This solution is valid provided that the conditions, $%
\epsilon _{S}/|\Delta +i\cdot \epsilon _{N}|\cdot \sqrt{\Delta /\Gamma }%
<<\{r_{S}^{2},r_{N}^{2}\}$ are fulfilled.

With the help of the functions $\{G^{R},\ F^{R}\}$, one can easily calculate
all the functions in the expression for the conductance $\delta S$ using
Eq.( \ref{eq2}). In calculating the conductance $G$ with the help of Eq.(2),
we used both analytical and numerical solutions of the Usadel equation, Eq.
(4). We calculate the temperature dependence of the conductance for both
types of structures shown in Fig.1. Our results show that the maximum near $%
T_{c}$ is not a specific characteristic of a coplanar geometry (Fig.1(B)),
but is present in both cases and related to a contribution of the subgap
current to the conductance. It is worth also mentioning that the position
and amplitude of the maximum do not depend essentialy on the length of the
structure (unless the Josephson effect becomes important in a S/N/S
structure). We distinguish the cases of long and short contacts in order to
make the theoretical analysis more transparent.

First we consider the structure shown in Fig1(A). It turns out that the
solution of the linearised Eq. (4) is a good approximation even in the case
of comparable interface and N film resistances ($r_{S}$ and $r_{N}$ are of
order 1). In Fig. 3 we plot the temperature dependence of the normalised
conductance variation $\delta S=(G-G_{n})/G_{n}$ for a long S/N/S structure
shown in Fig. 1(A) (here $G_{n}$ is the conductance of the system in the
normal state). As is seen from Fig. 3, a small maximum in $\delta S$ appears
at a temperature which is very close to $T_{c}$ and corresponds to the value
of $\Delta (T)$ comparable with the Thouless energy $\epsilon _{L}$. With
decreasing temperature the conductance variation decreases and becomes
negative as it should be if the interface resistance exceeds the resistance
of the N layer.

Using Eq.(5), we calculate the temperature dependence of the conductance for
the short structure shown in Fig. 1(A) for different values of parameters
and show this dependence in Fig.\ref{fig4} (two lower curves). As in the
previous case of a long structure there is a small maximum near $T_{c}$. In
this case the position of the maximum corresponds to $\Delta (T)$ equal
approximately to the energy $\epsilon _{S}$ at which the DOS in the N layer
has a maximum and below which the subgap conductance ($b$ term in Eq.(\ref
{eq2})) is enhanced. With decreasing temperature the conductance variation
decreases and becomes negative. The conductance variation $\delta S$ may
have a minimum and a second maximum at lower temperatures (subgap
conductance). Preliminary measurements carried out at low temperatures
reveal a minimum in the conductance in the temperature range 70 mK in
qualitative agreement with the theory. Analysing the energy dependence of
different functions in the intergrand of Eq.(2), one can see that both the
low temperature maximum and the maximum near $T_{c}$ are caused by a
contribution of the term $b(\epsilon )$ which describes the subgap charge
transfer through the S/N interface via Andreev reflection processes.

Consider now the coplanar structure (see Fig.\ref{fig1} (B)). One can show
that in this case the conductance variation is determined again by Eq.(\ref
{eq2}) with $r_{S}$ replaced by $\widetilde{r}_{S}=\sqrt{R_{S/N}d/\rho L^{2}}
$ and $(\nu _{S}\nu (L)+b(L))$ replaced by $\sqrt{m(\nu _{S}\nu +b)}$ (here $%
\rho /d$ is the sheet resistance of the semiconducting layer). In the case
of a short structure ($\widetilde{r}_{S}\gtrsim 1$) all the functions in the
expression for $\delta S$ ($\nu ,m$ etc) are constant in space and given by
Eq. (\ref{eq6}) with $\epsilon _{L}$ replaced by $\varepsilon _{d}\equiv
\hbar D/d^{2}$. This means that in a coplanar structure $\varepsilon _{d}$
plays the role of an effective Thouless energy. Therefore the position of
the maximum depends on $L$ weakly and is determined by the pseudogap $%
\varepsilon _{d}/2r_{s}$ induced in the N layer beneath the superconducting
electrode.

In Fig.\ref{fig4} we present the temperature dependence of the conductance
variation $\delta S$ near $T_{c}$ for the coplanar structure with parameters
close to the experimental ones. One can again see that there is a maximum
near $T_{c}$ at a temperature corresponding to $\Delta \cong 0.21\epsilon
_{d}$.

The presented theoretical results qualitatively fit the experimental data
and give a reasonable quantitative agreement. The normalized value of the
peak $\delta S_{\max }$ \ observed experimentally is about 10\% and its
position with respect to the critical temperature is $\delta T_{\max
}/T_{c}\cong 0.015.$ Theoretical values of $\delta S_{\max }$ vary from
tenth of percents (weak proximity effect) to 10\% (strong proximity effect)
and values of $\delta T_{\max }/T_{c}$ vary from $10^{-3}$ to $10^{-1}$.
Estimations of characteristic parameters for a coplanar structure give for $%
\widetilde{r}_{S},{r}_{S}$ and $\epsilon _{d}$ the approximate values 0.2-5,
5-10 and 5 K respectively. Therefore the theoretical values for $\delta
S_{\max }$ and $\delta T_{\max }/T_{c}$ are in reasonable agreement with
experimental data taking into account approximations made in the course of
calculations (for example, in case of S/N/S junctions, we neglect the
Josephson effect which seems to be suppressed by thermal fluctuations) and
uncertainties in the experimental value for the S/N interface resistance.

To summarise, we have for the first time observed a maximum in the
conductance of mesoscopic diffusive S/N, S/N/S structures in close vicinity
of the critical temperature. The maximum is caused by the contribution of
Andreev reflection processes to the pseudogap conductance of the S/N
interface. These processes are the ones which lead to the subgap conductance
at low temperatures. Our calculation shows that the conductance will peak
when the superconducting gap $\Delta (T)$ is of order of the subgap $%
\epsilon _{d}/2r_{S}$ which is in reasonable agreement with the experimental
observation. Studies of this maximum provide a new way to probe the effect
of the Coulomb interactions on the induced subgap in the semiconducting
region. The theory and experimental data presented here show that the
maximum in the conductance near $T_{c}$ is not related to non-stationary
Josephson effects.

We are grateful to D.E.Khmelnitskii, B.D.Simons and D. Taras-Semchuk for
discussions and encouragement at the initial part of the work and to V.
Pavlovksii for help in numerical calculations. We are grateful to the Royal
Society, the EPSRC and to the Isaac Newton Trust for their financial support.

\begin{figure}
\centerline{\psfig{figure=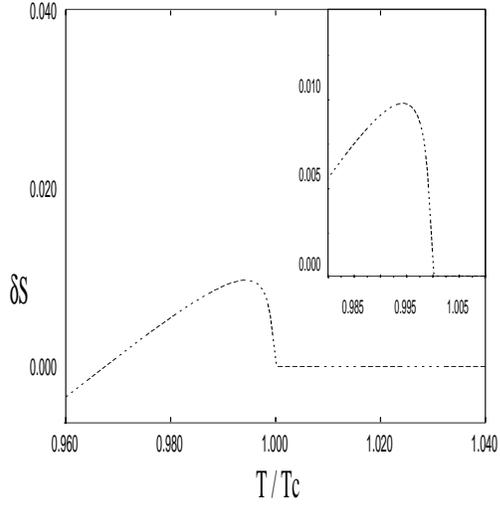,width=7cm,height=7cm}}
\caption{Normalised conductance variation $\protect\delta S$ vs normalised
temperature for a long S/N/S structure (Fig. 1(A)). The chosen parameters
are : $r_S=2,r_N=1,\;20 \protect\epsilon _L =\Delta (0),\;\Gamma =0.03\Delta
(0)$.}
\label{fig3}
\end{figure}

\begin{figure}
\centerline{\psfig{figure=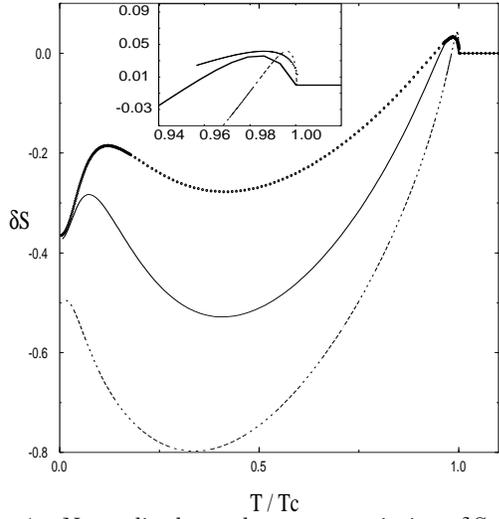,width=7cm,height=7cm}}
\caption{Normalised conductance variation $\protect\delta S$ vs normalised
temperature for a short N'/N/S structure (two lower curves) and coplanar
structure (upper curve). The parameters used are: $\Gamma =0.03\Delta
(0),r_{S}=r_{N}=5,\Delta (0)=2\protect\epsilon _{L}$ (dashed line); $\Gamma
=0.03\Delta (0),r_{S}=5,r_{N}=8.33,\Delta (0)=\protect\epsilon _{L}$ (solid
line); $\Gamma =0.03\Delta (0),{r_{S}}=5,r_{N}=8.33,\Delta (0)=\protect%
\epsilon _{d}$ (coplanar structure, dotted line). The maximum in the
conductance corresponds to a $\Delta =0.21 \protect\epsilon_{d}$.}
\label{fig4}
\end{figure}


\begin{references}
\bibitem{ftn1}  Strictly speaking, the position of this maximum, as we will
see, depends also on the interface resistance which makes an effective
Thouless energy smaller.
\bibitem{ftn3}  A portion of the conductance enhancement upon $T_{c}$ ($%
\approx 1-50\%$ for the various samples) is due to the vanishing resistance
of the aluminium layer between the voltage probes. In junction 1 (figure 3) $%
15\%$ of the conductance enhancment ($0.9mS$) upon the transition is due to
this effect.
\end{references}
\end{document}